\documentstyle[12pt,subeqn]{article}
\renewcommand{\theequation}{\arabic{section}.\arabic{equation}}

\begin{document}
\begin{flushright}
gr-qc/9407003
\end{flushright}
\begin{center}
\LARGE
Generation of Electro and Magneto Static Solutions \\
of the Scalar-Tensor Theories of Gravity\\
{}~\\~\\
\Large
William Bruckman\\
{}~\\
\normalsize
Penn State University, Department of Physics\\
104 Davey Laboratory, University Park, PA  16802\\
Telephone: (814) 863-3949
{}~\\~\\
\end{center}
\begin{abstract}
The field equations of the scalar-tensor theories of gravitation are
presented in different representations, related to each other by
conformal transformations of the metric.  One of the representations
resembles the Jordan-Brans-Dicke theory, and is the starting point for
the generation of exact electrostatic and magnetostatic exterior
solutions.  The corresponding solutions for each specific theory can be
obtained by transforming back to the original canonical representation,
and the conversions are given for the theories of Jordan-Brans-Dicke,
Barker, Schwinger, and conformally invariant coupling.  The
electrostatic solutions represent the exterior metrics and fields of
configurations where the gravitational and electric equipotential
surfaces have the same symmetry.  A particular family of electrostatic
solutions is developed, which includes as special case the spherically
symmetric solutions of the scalar-tensor theories.  As expected, they
reduce to the well-known Reissner-Nordstr\"{o}m metric when the scalar
field is set equal to a constant.  The analysis of the
Jordan-Brans-Dicke metric yields an upper bound for the mass-radius
ratio of static stars, for a class of interior structures.
{}~\\~\\
PACS: 04.20Jb\\
{}~\\
On sabbatical leave from  University of Puerto Rico at Humacao,
Department of Physics.
\end{abstract}

\newpage
\addtolength{\baselineskip}{\baselineskip}
\section{Introduction and Summary}
\setcounter{equation}{0}

	The scalar-tensor theories of gravitation$^{1-4}$
differ from Einstein's general relativity since they admit the existence
of a long range scalar field coupled to matter in such a way that the
theories satisfy the weak principle of equivalence, and therefore are
viable alternatives for the explanation of gravitational phenomena.
However, the strong principle of equivalence is not valid, since the
scalar field connects the local physics with the mass-energy of the rest
of the universe.  In fact, the Jordan-Brans-Dicke[JBD] theory, which has
the simplest scalar coupling, was formulated as a relativistic model
allowing a better representation of Mach's principle and Dirac's ideas
concerning the connection between the Newtonian gravitational constant
and the age of the universe$^{5}$.  The roots of the scalar-tensor
theories go back to the original Kaluza-Klein five dimensional
unification of the gravitational and electromagnetic forces, and the
modifications by Einstein-Mayer$^{6}$, Jordan$^{7}$, and
Thiry$^{8}$.  The five dimensional metric of these theories has five
additional degrees of freedom, where four of
them serve to represent the electromagnetic vector potential while the
other behaves as a scalar field in a four-dimensional geometry.  Thus we
have that a scalar field appears as a natural outcome of the intriguing
Kaluza-Klein unification and, furthermore, is the simplest long range
field by means of which the matter distribution of the universe can
affect local physics.  The scalar-tensor formalism is able to
incorporate this hypothetical but interesting interaction consistently
and in agreement with experiments, and hence opens a broader theoretical
framework to approach the study of gravitation.

	Aside from the physical and philosophical motivations for
considering the existence of scalar fields, they could also provide
heuristic representations of matter fields which, due to their
mathematical simplicity, help us gain physical insights.  For instance,
it is well known that an isentropic fluid with pressure equal to
energy density can be represented by a scalar field.  Moreover, they
could also be interpreted as large perturbations, serving in the study
of the stability of general relativity backgrounds.

	There are several alternative representations of the
scalar-tensor theories which are related to the original canonical
representation, Eqs. (2.1) and (2.2), by a conformal transformation of
the metric, where the conformal factor is a function of the scalar
field.  For example, in the Einstein-scalar representation the new
metric, $\overline{g}_{\mu \nu}$, and the original metric,
$g_{\mu \nu}$, are related by

\begin{displaymath}
\overline{g}_{\mu \nu} = \phi g_{\mu \nu}.
\end{displaymath}
The field equations for $\overline{g}_{\mu \nu}$, given by Eqs. (2.6) and
(2.7), are as in Einstein's theory but with an added energy-momentum
tensor for the scalar field.   A more general transformation is
\begin{displaymath}
\tilde{g}_{\mu \nu} = \frac{\phi}{\tilde{\phi}} g_{\mu \nu},
\end{displaymath}
where the new scalar field $\tilde{\phi}$ is an arbitrary function of
$\phi$, and we arrived at the field equations (2.12) and 2.13) which are
remarkably similar to the originals.  Alternatively, we can reexpress
the Einstein-scalar representation using now $\tilde{\phi}$ instead of
$\phi$ as the scalar field to obtain Eqs. (2.17) and (2.18).  A
particular choice of the function $\tilde{\phi}(\phi)$ makes the theory
look similar to the JBD theory in the sense that the second order
equations for the metric, Eqs. (2.12) and (2.17), take the same form as
the corresponding one in the JBD theory, with a constant parameter,
$\omega _{0}$, playing the role of the scalar field coupling constant.
When $\omega _{o}$ is chosen to be zero the theory can be written as a
five dimensional Einstein's field equations in which the geometry is
independent of the fifth dimension (Eqs. (2.22) - (2.28)).

	The representation $\overline{g}_{\mu \nu}, \tilde{\phi}$, is
suitable for the mathematical analysis, and therefore is the starting
point in Section III for the generation of exact scalar-tensor
electrostatic solutions, starting from given static vacuum metrics in
general relativity.  This mapping gives the exterior fields of
configurations of arbitrary shape, for which the surfaces of constant
electrostatic and gravitational potential coincide.  They can be converted
into new solutions
of the original canonical representation, $\phi, g_{\mu \nu}$, for each
particular theory, and the necessary transformations are then given
for the specific theories of JBD, Barker, Schwinger, and for
conformally invariant coupling.  Solutions representing the exterior
gravitational, electric, and scalar fields of oblates or prolates
configurations are presented explicitly in Section IV.  These solutions
reduce to the Reissner-Nordstr$\ddot{o}$m metric when the scalar field is set
equal to a constant and the configuration is taken to be spherically
symmetric.

	The spherically symmetric solutions are study in Section V,
where it is found that, in contrast to general relativity, in the
scalar-tensor theories the analysis
of the exterior solution alone is not sufficient to establish and upper
bound for the mass-radius ratio of the static configurations, and
it is necessary to consider the interior configurations as well.
This is done in detail for the JBD theory, where an upper bound is
determined for equations of state where the volume integral of the trace
of the energy-momentum tensor is positive.

	In the last section, exact scalar-tensor magnetostatic solutions
are given in terms of vacuum metrics in general relativity.  Then an
explicit family of solutions is worked out, that represents prolates or
oblates configurations immersed in magnetic and scalar fields.
These solutions are scalar-tensor
theories generalizations of Melvin's magnetized metric of general
relativity.

\section{The Field Equations}
\setcounter{equation}{0}

	The scalar-tensor theories field equations, without a
cosmological term, can be written in the following canonical form:
\begin{equation}
\phi G_{\mu \nu} = 8 \pi GT_{\mu \nu} + \frac{\omega}{\phi} \left [
\phi_{,\mu}\phi_{,\nu} - \frac{1}{2} g_{\mu \nu} \phi^{, \alpha}
\phi_{,\alpha} \right ] + \phi_{; \mu \nu} - g_{\mu \nu} \phi^{;
\alpha}_{; \alpha},
\end{equation}
\begin{equation}
(2 \omega + 3) \phi^{; \alpha}_{; \alpha} = \frac {2 \omega +
3}{\sqrt{-g}} \left [ \sqrt{-g} \phi^{,\alpha} \right ]_{, \alpha} = 8
\pi GT^{\alpha}_{\alpha} - \phi^{, \alpha} \phi_{, \alpha} \frac{d
\omega}{d \phi} ,
\end{equation}
where, as usual, commas and semicolons mean partial and covariant
derivatives, respectively, $G_{\mu \nu}$ is the Einstein tensor, and
$T_{\mu \nu}$ is the energy momentum tensor for all fields excluding the
scalar, $\phi$, and the metric, $g_{\mu \nu}$, fields.  We are
essentially adopting the notation convention of Misner-Thorne and
Wheeler$^{9}$.  It follows from Eqs. (2.1), (2.2) and the Bianchi
identities that we still have, as it is in general relativity, the
energy-momentum conservation:
\begin{equation}
T^{\mu \nu}\,_{; \nu} = 0.
\end{equation}
Thus, in this representation, the scalar field does not enter the
equation of motion, and consequently free falling test particles move in
geodesics of the metric $g_{\mu \nu}$.  With the simplest choice of
coupling $\omega =const$, we get the well known JBD field equations.
Other examples of interesting theories can be defined by
\begin{equation}
\omega = \frac{1}{2} \left[ \frac{k_{1}}{k_{2} \phi + k_{3}} -3 \right ] ,
\end{equation}
where $k_{1}, k_{2}$ and $k_{3}$ are constants.  For instance, if
$k_{1}=k_{2}=-k_{3}=1$, we have the theory due to Barker$^{10}$, while
the case $k_{3}=0$ was motivated by Schwinger$^{11}$.  Another attractive
model is the conformally invariant curvature coupling$^{1}$, in which
$k_{1}=3, k_{2}=-1$, and $k_{3}=1$.

	An alternative representation of these theories is provided by
the conformal transformation$^{12}$
\begin{equation}
\overline{g}_{\mu \nu} = \phi g_{\mu \nu},
\end{equation}
which gives rise to the Einstein's scalar form of the field equations:
\begin{equation}
\overline{G}_{\mu \nu}=8\pi \overline{T}_{\mu \nu} + \left ( \omega +
\frac{3}{2} \right ) \left[ \Phi_{,\mu} \Phi_{,\nu} - \frac{1}{2}
\overline g_{\mu \nu}
\Phi^{, \alpha} \Phi_{, \alpha} \right ] ,
\end{equation}
\begin{equation}
\frac{2 \omega +3}{\sqrt{- \overline{g}}} \left [ \sqrt{- \overline{g}}
\Phi^{, \alpha} \right ]_{, \alpha} = 8 \pi
\overline{T}^{\alpha}_{\alpha}-\Phi^{, \alpha} \Phi_{, \alpha} \frac{d
\omega}{d \Phi},
\end{equation}
where
\begin{equation}
\Phi \equiv \ell n \phi ,
\end{equation}
and
\begin{equation}
\overline{T}_{\mu \nu} = \frac{T_{\mu \nu}}{\phi}  .
\end{equation}
The bar over $G_{\mu \nu}$ means that $\overline{g}_{\mu \nu}$ is
replacing $g_{\mu \nu}$, and now indices are raised with
$\overline{g}_{\mu \nu}$.

	A more general transformation is generated by letting
\begin{equation}
\tilde{g}_{\mu \nu} = \frac{\phi g_{\mu \nu}}{\tilde{\phi}} =
\frac{\overline{g}_{\mu \nu}}{\tilde{\phi}},
\end{equation}
\begin{equation}
\tilde{\phi} = \tilde{\phi}(\phi),
\end{equation}
where $\tilde{\phi}$ is an arbitrary function of $\phi$.  The new field
equations can be written in the form
\begin{equation}
\tilde{\phi} \tilde{G}_{\mu \nu} = 8 \pi G \tilde{T}_{\mu \nu} +
\frac{\tilde{\omega}}{\tilde{\phi}} \left [ \tilde{\phi}_{, \mu} \tilde
{\phi}_{, \nu} - \frac{1}{2} \tilde{g}_{\mu \nu} \tilde{\phi}^{, \alpha}
\tilde{\phi}_{, \alpha} \right ] + \tilde{\phi}_{; \mu \nu} - \tilde{g}_{\mu
\nu}
\tilde{\phi}^{; \alpha}_{; \alpha} ,
\end{equation}
\begin{equation}
\frac{2 \tilde{\omega} + 3}{\sqrt{- \tilde{g}}} \left[ \sqrt{- \tilde{g}}
\tilde{\phi}^{, \alpha} \right ]_{, \alpha} = 8 \pi \frac{G
\tilde{T}^{\alpha}_{\alpha}}{\sigma} - \tilde{\phi}^{, \alpha}
\tilde{\phi}_{, \alpha} \frac{d \tilde{\omega}}{d \tilde{\phi}} ,
\end{equation}
where
\begin{equation}
\frac{\tilde{T}_{\mu \nu}}{\tilde{\phi}} = \frac{T_{\mu \nu}}{\phi}
\end{equation}
\begin{equation}
\sigma \equiv \frac{d \ell n \tilde{\phi}}{d \ell n \phi}
\end{equation}
\begin{equation}
2 \tilde{\omega} + 3 \equiv \frac{2 \omega + 3}{\sigma^{2}}
\end{equation}
Yet, if instead of $\tilde{g}_{\mu \nu}$, the Einstein-scalar metric
$\overline{g}_{\mu \nu}$ is used, we will obtain
\begin{equation}
\overline{G}_{\mu \nu}=8\pi G \overline{T}_{\mu \nu} + \left (
\tilde{\omega} + \frac{3}{2} \right ) \left [ \tilde{\Phi}_{, \mu}
\tilde{\Phi}_{, \nu} - \frac{1}{2} \overline{g}_{\mu \nu}
\tilde{\Phi}^{, \alpha} \tilde{\Phi}_{, \alpha} \right ]
\end{equation}
\begin{equation}
\frac{2 \tilde{\omega}+3}{\sqrt{-\overline{g}}} \left [ \sqrt{- \overline{g}}
\tilde{\Phi}^{, \alpha} \right ]_{, \alpha} = 8 \pi \frac{G
\overline{T}^{\alpha}_{\alpha}}{\sigma} - \tilde{\Phi}^{, \alpha}
\tilde{\Phi}_{, \alpha} \frac{d \tilde{\omega}}{d \tilde{\Phi}}
\end{equation}
where
\begin{equation}
\tilde{\Phi} \equiv \ell n \tilde{\phi}.
\end{equation}
An interesting choice for the new scalar field is
\begin{equation}
\tilde{\Phi} = \frac{1}{\sqrt{2\omega_{0}+3} } \int \sqrt{2 \omega + 3}
d\Phi + \;\hbox{Const}.
\end{equation}
\begin{equation}
\omega_{0} = \;\hbox{Const};
\end{equation}
since it then can be shown from Eqs. (2.15) and (2.16) that
$\tilde{\omega}=\omega_{0}$, and therefore the field equations take the
JBD form, except for the factor $\frac{1}{\sigma}$ multiplying
$\tilde{T}^{\alpha}_{\alpha}$ and $\overline{T}^{\alpha}_{\alpha}$ in
Eqs. (2.13) and (2.18).  Nevertheless, here $\omega_{0}$ is an arbitrary
parameter, which, of course, for the JBD theory can be set equal to
$\omega$.

	The peculiar value $\omega_{0}=0$ yields a representation that
can be cast in the following five-dimensional language:
\begin{equation}
\stackrel{5}{g}_{\mu \nu} = \tilde{g}_{\mu \nu}\; ; \; \mu,\nu=\{1-4\},
\end{equation}
\begin{equation}
\stackrel{5}{g}_{\mu 5}=0
\end{equation}
\begin{equation}
\stackrel{5}{g}_{5 5}=\tilde{\phi}^{2}
\end{equation}
\begin{equation}
\stackrel{5}{G} \stackrel{5}{T}_{\mu \nu}=\frac{G \tilde{T}_{\mu
\nu}}{\tilde{\phi}}\; ; \; \stackrel{5}{G} = \;\hbox{const.},
\end{equation}
\begin{equation}
\stackrel{5}{G} \stackrel{5}{T}_{\mu 5} =  0
\end{equation}
\begin{equation}
\stackrel{5}{G} \stackrel{5}{T}_{55} = \frac{1}{2} \tilde{\phi}
\left [ 1-\frac{1}{\sigma}\right ] G \tilde{T}^{\alpha}_{\alpha},
\end{equation}
and hence the scalar tensor field equations become
\begin{equation}
\stackrel{5}{G}_{AB} = 8\pi \stackrel{5}{G} \stackrel{5}{T}_{AB} \; ; \;
A, B=\{1-5\},
\end{equation}
where $\stackrel{5}{G}_{AB}$ is the five-dimensional Einstein's tensor
for the metric $\stackrel{5}{g}_{AB}(X^{\mu})$.

	We can readily integrate Eq. (2.20) for the  $\omega$ given by
Eq. (2.4), and we obtain:

\noindent
(i) {\em Barker's Theory}
\begin{equation}
\ell n \tilde{\phi} = \frac{2}{\ell} Tan^{-1} \sqrt{\phi -1} +\;
\hbox{Const.}
\;;\; \ell \equiv \sqrt{2 \omega_{0}+3},
\end{equation}
\begin{equation}
\phi = Tan^{2} \frac{\ell}{2} [\ell n \tilde{\phi} -\; \hbox{Const.}] + 1 .
\end{equation}
\noindent
(ii) {\em Schwinger's Theory}
\begin{equation}
\ell n \tilde{\phi} = \frac{-2}{\ell \sqrt{\alpha \phi}} +\;
\hbox{Const.}\;;\;
\alpha \equiv \frac{k_{2}}{k_{1}}
\end{equation}
\begin{equation}
\phi = \frac{4}{\alpha \ell^{2}} [\; \hbox{Const.} - \ell n
\tilde{\phi}]^{-2} .
\end{equation}
\noindent
(iii) {\em Conformally Invariant Curvature Coupling}
\begin{equation}
\ell n \tilde{\phi} = \frac{\sqrt{3}}{\ell} \ell n \left [
\frac{1-\tau}{1+\tau} \right ] + \; \hbox{Const.}, \tau \equiv \sqrt{1-\phi}
\end{equation}
\begin{equation}
\phi = \frac{4 \hbox{Const}' \tilde{\phi}^{\ell/ \sqrt{3}}}
{\left [ \tilde{\phi}
^{\ell/ \sqrt{3}} + \; \hbox{Const.}' \right ]^{2}}
\end{equation}
\section{Generation of Electrostatic Solutions}
\setcounter{equation}{0}

	Let's consider the field equations (2.17) and (2.18) in the
representation where $\tilde{\omega}=\omega_{0}$ [JBD like representation], and
assume that we have the energy-momentum tensor of an electromagnetic
field:
\begin{equation}
8 \pi G \overline{T}_{\mu \nu} = 2 \left[ F_{\mu \alpha}
F_{\nu}^{\alpha} - \frac{1}{4} \overline{g}_{\mu \nu} F^{2} \right ]
\end{equation}
\begin{equation}
F_{\mu \nu} = \frac{\partial A_{\nu}}{\partial X^{\mu}} - \frac{\partial
A_{\mu}}{\partial X^{\nu}} .
\end{equation}
	Furthermore, let's assume that the space-time is static, and the
vector potential only has the component $A_{4}$.  Hence, we are dealing
with three potentials; the electrostatic potential $A_{4}$, and the
gravitational potentials which are represented by $\overline{g}_{44}$
and $\tilde{\phi}$. We know that when the configurations are spherically
symmetric these potentials depend only on the radius variable that
characterize the equipotential surfaces.  Moreover, even if the
configurations are not spherically symmetric, we could still expect that
the gravitational and electric equipotentials are surfaces of the same
shape, and, consequently, these potentials once more would be identified
by a single function, say $u$.  Thus, in searching for new solutions, it
is quite natural to start with the following ansatz:
\begin{equation}
A_{4} = A_{4}(u),
\end{equation}
\begin{equation}
\tilde{\phi} = \tilde{\phi}(u),
\end{equation}
\begin{equation}
\overline{g}_{44} = \overline{g}_{44}(u).
\end{equation}

	Then one finds (Appendix A) that the electrostatic solutions of the
JBD-like field equations are expressed in terms of solutions, $g^{'}_{\mu
\nu}$, of Einstein's static vacuum field equations, $G^{'}_{\mu \nu}=0$,
in the following way:
\begin{equation}
- \overline{g}_{44} = \frac{4ab(-g'_{44})^{c}}{[b-(-g'_{44})^{c}]^{2}} ,
\end{equation}
\begin{equation}
\overline{g}_{ij} = \frac{[b-(-g'_{44})^{c}]^{2}}{4ab} (-g'_{44})^{1-c}
g'_{ij} ; i, j=1,2,3,
\end{equation}
\begin{equation}
A_{4}= \frac{\pm \sqrt{a}[b+(-g'_{44})^{c}]}{b-(-g'_{44})^{c}} +d,
\end{equation}
\begin{equation}
\tilde{\Phi} = e\;\; \ell n(-g'_{44}) + \; \hbox{Const.},
\end{equation}
\begin{equation}
e \equiv \pm \sqrt{\frac{1-c^{2}}{2 \omega_{0}+3}} \; ; \; c^{2} < 1,
\end{equation}
where a, b, c, and d are constants.

	For asymptotically Minkowski space-time where

\begin{equation}
-g'_{44} \approx 1-\frac{2GM'}{r} \; ; \; r= \sqrt{X_{i}X^{i}}
\rightarrow \infty,
\end{equation}
we have that, with the choice
\begin{equation}
a=\frac{(b-1)^{2}}{4b},
\end{equation}
\begin{equation}
d= \pm \sqrt{a} \frac{[1+b]}{[1-b]} ,
\end{equation}
the asymptotic behavior of $\overline{g}_{\mu \nu}, \tilde{\phi}$ and
$A_{4}$ are
\begin{equation}
- \overline{g}_{44} \approx 1 - \frac{2GM}{r}
\end{equation}
\begin{equation}
\overline{g}_{ij} \approx g'_{ij} \left [ 1+\frac{2GM'}{r} \left ( c-1+
\frac{2c}{b-1} \right ) \right ]  ,
\end{equation}
\begin{equation}
\tilde{\phi} \approx const. \left[1 - \frac{2GeM'}{r} \right] ,
\end{equation}
\begin{equation}
A_{4} \approx - \frac{kQ}{r} ,
\end{equation}
where $k$ is a constant, and
\begin{equation}
M = cM' \left ( \frac{b+1}{b-1} \right )
\end{equation}
\begin{equation}
kQ = \pm \frac{2c \sqrt{b} GM'}{b-1}
\end{equation}
\begin{equation}
GM^{2} = (GcM')^{2} + (kQ)^{2}
\end{equation}
Note that if $b \rightarrow \infty$, then $Q \rightarrow 0$, and
$M=cM'$.  Thus for $M$ and $M'$ to be positive we must assume
$c>0$.  From Eqs. (3.14) and (3.17) we see that $M$ and $Q$ can be identified
with the total inertial mass and electric charge, respectively.

	The solutions (3.6) - (3.9) can be converted into the
corresponding solutions of the original scalar-tensor field equations
(2.1) and (2.2) by using the transformation (2.10):
\begin{equation}
g_{\mu \nu} = \frac{\overline{g}_{\mu \nu}}{\phi}  ,
\end{equation}
and then $\phi(\tilde{\phi})$ could be determined by the inversion of
Eq. (2.20), with $\tilde{\phi}$ given, by Eq. (3.9), explicitly as
a function of $g'_{44}$.  For instance, for the JBD theory we can have
$\phi = \tilde{\phi}$, while for Barker's Schwinger's and conformally
invariant curvature coupling theories, the necessary inversions are
found in Eqs. (2.30), (2.32),and (2.34), respectively.

\section{Weyl Type Solutions}
\setcounter{equation}{0}

	An interesting family of solutions are generated with the
transformation (3.6)-(3.9) when we take as seed metric, $g'_{\mu \nu}$,
the following well known Weyl solution:
\begin{equation}
-g'_{44} = \left [ 1-\frac{2 \beta}{r} \right ]^{\delta}
\end{equation}
\begin{equation}
g'_{rr}= \frac{\left [ 1-\frac{2 \beta}{r} \right
]^{\delta^{2}-\delta-1}}{\left [ 1-\frac{2 \beta}{r} + \frac{\beta
2}{r^{2}}
sin^{2} \theta \right ] ^{\delta^{2}-1}}
\end{equation}
\begin{equation}
g'_{\theta \theta} = \frac{r^{2}\left [ 1-\frac{2 \beta}{r} \right ]
^{\delta^{2}-\delta}}{\left [ 1-\frac{2 \beta}{r} + \frac{\beta^{2}}{r^{2}}
sin^{2} \theta \right ] ^{\delta^{2}-1}}
\end{equation}
\begin{equation}
g'_{\varphi \varphi} = r^{2}\left [1-\frac{2 \beta}{r} \right ]^{1-\delta}
sin^{2} \theta
\end{equation}
where $\beta$ and $\delta$ are constants.  Clearly, the above solutions
reduce to the Schwarzschild metric when $\delta =1$.  Otherwise, they
represent the gravitational field exterior to oblates $(\delta >1)$ or
prolates $(\delta < 1)$ ellipsoidal configurations$^{13}$.

	The new solutions generated by Eq. (3.6)-(3.9) are:
\begin{equation}
- \overline{g}_{44} = \left ( \frac{b-1}{b-\left[ 1-\frac{2 \beta}{r}
\right ]^{\delta c}} \right )^{2} \left [ 1- \frac{2 \beta}{r} \right
]^{\delta c}
\equiv \psi \left [1 - \frac{2 \beta}{r} \right ]^{\delta c}
\end{equation}
\begin{equation}
\overline{g}_{rr} = \frac{\left[ 1 - \frac{2 \beta}{r} \right
]^{\delta^{2}-\delta c-1}}{\psi \left[1 - \frac{2 \beta}{r} +
\frac{\beta^{2}}{r^{2}} sin^{2} \theta \right ]^{\delta^{2}-1}}
\end{equation}
\begin{equation}
\overline{g}_{\theta \theta} = \frac{r^{2}[1-\frac{2
\beta}{r}]^{\delta^{2}-\delta c}}{\psi [1- \frac{2 \beta}{r} +
\frac{\beta^{2}}{r^{2}} sin^{2} \theta]^{\delta^{2}-1}}
\end{equation}
\begin{equation}
\overline{g}_{\varphi \varphi} = \frac{r^{2}[1-\frac{2 \beta}{r}]^{1-\delta
c}}{\psi}
sin^{2}\theta
\end{equation}
\begin{equation}
A_{4}=\pm \left [ \frac{b-1}{2 \sqrt{b}} \left ( \frac{b+[1-\frac{2
\beta}{r}]^{\delta c}}{b-[1-\frac{2 \beta}{r}]^{\delta c}} \right ) -
\frac{b+1}{2 \sqrt{b}} \right ]
\end{equation}
\begin{equation}\tilde {\Phi} = \delta e\;\;\ell n \left[ 1 - \frac{2 \beta}{r}
\right ] +\;\hbox{Const.}
\end{equation}
where
\begin{equation}
\psi \equiv \left ( \frac{b-1}{b-[1-\frac{2 \beta}{r}]^{\delta c}}
\right )^{2}
\end{equation}
The solutions (4.5)-(4.10) represent static gravitational, electric, and
scalar fields with oblate or prolate equipotential surfaces (Appendix
B).  When $\delta = 1$ the solutions are spherically symmetric, and the
metric takes this simple form:
\begin{equation}
\overline{g}_{\theta \theta} = \frac{\overline{g}_{\varphi \varphi}}{
sin^{2} \theta}= \left( \frac{b-[1-\frac{2 \beta}{r}]^{c}}{b-1} \right
)^{2} \left [ 1-\frac{2 \beta}{r} \right]^{1-c} r^{2} \equiv
\overline{R}^{2}
\end{equation}
\begin{equation}
- \overline{g}_{44} = \overline{g}_{rr}^{-1} = \frac{(b-1)^{2}[1- \frac{2
\beta}{r}]^{c}}{(b-[1-\frac{2 \beta}{r}]^{c})^{2}}
\end{equation}
The above solution is a scalar-tensor generalization of the
Reissner-Nordstr$\ddot{o}$m metric, to which it reduces when $c=1$ (Appendix
C).

	If $c \neq$ 1, and $r=2 \beta = 2GM'$, we see from Eq. (4.13) that
$g'_{44}=0$.  However, in contrast to general relativity where $r=2GM'$
represents an event horizon, here we have, from Eq. (4.12), that
\begin{displaymath}
\overline{R} \rightarrow 0,
\end{displaymath}
\begin{equation}
\hbox{lim} \;\;r \rightarrow 2 \beta
\end{equation}
and, since the proper areas of constant radius of the new solutions are
equal to $4 \pi \overline{R}^{2}$, then we see that the Schwarzschild
horizon is mapped into the origin of the JBD like representation.  Thus,
we have that the static metric (4.12), (4.13) and scalar field (4.10)
describe the whole space $0< \overline{R} < \infty$, for $2GM' < r <
\infty$, without a horizon.  Furthermore, this exterior metric does not
imply a lower bound for $\overline{R}$, analogous to $r>2GM'$ appearing
in general relativity.  Nevertheless, we have to be cautious with the
physical interpretation of $\overline{R}$, since now we are working in
the JBD like representation where the physical meaning of the metric and
scalar field $\tilde{\phi}$ are not transparent, for we have that free
falling test particles do not move in geodesics of the metric
$\overline{g}_{\mu \nu}$.  Instead, we should look at the behavior of
the radius, $R(r)$, in the original canonical representation, $g_{\mu
\nu}, \phi$, which, by virtue of Eq. (2.5), is related to $\overline{R}$
in the following way
\begin{equation}
R = \sqrt{g_{\theta \theta}} = \sqrt{\frac {\overline{g}_{\theta
\theta}}{\phi}} = \frac{\overline{R}}{\sqrt{\phi}}.
\end{equation}
In the next section we will take on this analysis for the JBD theory.

\section{Analysis Of R(r).}
\setcounter{equation}{0}

	For the JBD theory we can have $\omega = \omega_{0}, \phi =
\tilde{\phi}$, and hence from Eq. (4.10) we find that
\begin{equation}
\phi = \left [ \frac{1-{2GM'}}{r} \right ]^{e}, \delta = 1,
\end{equation}
where the constant of integration has been chosen such that $\phi \rightarrow
1$ as $r \rightarrow \infty$.  Using Eqs. (4.12), (4.15) and (5.1), we
get
\begin{equation}
R = r \left [ \frac{b-(1-\frac{2GM'}{r})^{c}}{b-1} \right ] \left [
1-\frac{2GM'}{r} \right ]^{\frac{1-c-e}{2}},
\end{equation}
and depending on the sign of $S \equiv \frac{1-c-e}{2}$ we will have
as $r \rightarrow 2GM'$ that either $R \rightarrow 0 (S>0)$, or $R
\rightarrow \infty (S<0)$ after reaching a minimum value for some $r >
2GM'$.  On the other hand, the sign of $S$ is determined by the value of
$c$, which, in turn, depends on the interior structure of the star, as
described below.

	The field equations (2.17) and (2.18) with
$\tilde{\omega}=\omega_{0}$ imply for a static system that
\begin{equation}
\left[ \sqrt{- \overline{g}}(\ell n - \overline{g}_{44})^{,i}
\right]_{,i} = -16 \pi G \left[ \overline{T}^{4}_{4} - \frac{1}{2}
\overline{T}^{\alpha}_{\alpha} \right] \sqrt{- \overline{g}},
\end{equation}
\begin{equation}
\left [ \sqrt{- \overline{g}} \tilde{\Phi}^{,i} \right]_{,i} = \frac{8
\pi G \overline{T}^{\alpha}_{\alpha}}{\sigma(2 \omega_{0} + 3)}\sqrt{-
\overline{g}},
\end{equation}
and consequently
\begin{equation}
\int \sqrt{- \overline{g}}(\ell n - \overline{g}_{44})^{,i} d^{2}S_{i} =
- 16 \pi G \int \left [ \overline{T}_{4}^{4} - \frac{1}{2}
\overline{T}^{\alpha}_{\alpha} \right ] \sqrt{- \overline{g}} d^{3}X.
\end{equation}
\begin{equation}
\int \sqrt{- \overline{g}} \tilde{\Phi}^{,i} d^{2}S_{i} = 8 \pi G \int
\frac{\overline{T}^{\alpha}_{\alpha} \sqrt{-
\overline{g}}d^{3}X}{\sigma(2 \omega_{0}+3)}.
\end{equation}
\noindent
Thus, if the two dimensional surface of integration, $d^{2}S_{i}$, is
chosen outside the boundary of the star, we can then substitute, in the
left hand side of Eqs. (5.5) and (5.6), for $\overline{g}_{44}$ and
$\tilde {\Phi}$ the expressions given in Eqs. (3.6) and (3.9).  At this
point we simplified the problem by assuming that the configuration is
neutral $(Q=0)$, and consequently Eqs. (5.5) and (5.6) become
\begin{equation}
c \int \sqrt{- \overline{g}}( \ell n - g'_{44})^{,i} d^{2}S_{i} = - 16
\pi G \int \left [ \overline{T}^{4}_{4} - \frac{1}{2} \overline
{T}^{\alpha}_{\alpha} \right] \sqrt{- \overline{g}}d^{3}X,
\end{equation}
\begin{equation}
e \int \sqrt{-\overline{g}}(\ell n -g'_{44})^{,i} d^{2}S_{i} = 8 \pi G
\int \frac{\overline{T}^{\alpha}_{\alpha}\sqrt{- \overline{g}}d^{3}X}
{\sigma(2 \omega_{0}+3)}.
\end{equation}
{}From which it follows that
\begin{equation}
\frac{e}{c} = \frac{\int \frac{\overline{T}^{\alpha}_{\alpha}\sqrt{-
\overline{g}}d^{3}X}{\sigma (2 \omega_{0}+3)}}{-2 \int \left[
\overline{T}^{4}_{4} - \frac{1}{2}
\overline{T}^{\alpha}_{\alpha} \right] \sqrt{- \overline{g}} d^{3}X}=
\frac{\int
\frac{T^{\alpha}_{\alpha} \sqrt{-g}d^{3}X}{\sigma (2 \omega_{0}+3)}}{-2
\int \left[ T^{4}_{4} - \frac{1}{2} T^{\alpha}_{\alpha}\right ]
\sqrt{-g}d^{3}X},
\end{equation}
where we used Eqs. (2.5) and(2.9).  For a perfect fluid we obtain
\begin{equation}
\frac{e}{c}= \frac{\int\frac{(3p- \rho)\sqrt{-g}d^{3}X}{\sigma(2
\omega_{0}+3)}}{\int(3p+ \rho) \sqrt{-g} d^{3}X}.
\end{equation}
In particular, if the pressure, $p$, is proportional to the energy
density, $p=\epsilon \rho$, we find
\begin{equation}
\frac{e}{c} = \frac{(3 \epsilon -1)}{(3 \epsilon + 1)\sqrt{2
\omega_{0}+3}} \overline{\left ( \frac{1}{\sqrt{2 \omega +3}}\right )},
\end{equation}
where we use Eq. (2.16) with $\tilde{\omega}=\omega_{0}$, and
\begin{equation}
\overline{\left ( \frac{1}{\sqrt{2 \omega +3}} \right )} \equiv \frac{\int
\frac{\rho \sqrt{-g}d^{3}X}{\sqrt{2 \omega+3}}}{\int \rho
\sqrt{-g}d^{3}X}.
\end{equation}
Hence for the JBD theory we get, with $\omega_{o} = \omega$,
\begin{equation}
\frac{e}{c} = \frac{3 \epsilon -1}{(2 \omega +3)(3 \epsilon +1)},
\end{equation}
from which, using Eq. (3.10),
\begin{equation}
c = \sqrt{\frac{2 \omega +3}{2 \omega +3+ \left (\frac{3 \epsilon -1}{3
\epsilon + 1} \right )^{2}}}.
\end{equation}
If $\epsilon$ is not a constant, or we do not have a perfect fluid we
can still use Eqs. (5.13) and (5.14) by replacing $\epsilon$ in them by
\begin{equation}
\overline {\epsilon} \equiv \frac{- \frac{1}{3} \int T^{i}_{i} \sqrt{-g}
d^{3}X}{\int T^{4}_{4} \sqrt{-g}d^{3}X}= \frac{1}{3} \left [ \frac{-
\int T^{\alpha}_{\alpha} \sqrt{-g} d^{3}X}{\int T^{4}_{4} \sqrt{-g}
d^{3}X}+1 \right ].
\end{equation}
Using Eq. (5.13) and (5.14), we can express  $S$ as a function of
$\omega$ and $\overline{\epsilon}$:
\begin{equation}
S = \frac{1}{2}(1-c-e) = \frac{1}{2}
\left ( 1 - \sqrt{
        \frac{2 \omega+3}{2 \omega+3+\left(
            \frac{3\overline{\epsilon}-1}{3 \overline{\epsilon}+1}
                  \right)^{2}}}\left[ 1+
        \frac{3 \overline{\epsilon}-1}
        {(2 \omega +3)(3\overline{\epsilon}+1)}\right]
\right).
\end{equation}
Thus, assuming that $2 \omega +3>0$, if $0 \leq \overline{\epsilon} <
\frac{1}{3}$, then $S > 0$, and when $\overline{\epsilon} >
\frac{1}{3}$, one finds that $S < 0$.  The case
$\overline{\epsilon}=\frac{1}{3}$ yields $c=1$, which implies that $\phi
= const$, and we are back to the Schwarzschild metric of general
relativity.  According to Eq. (5.15), for a positive energy density,
$-T^{4}_{4} > 0$, we have $\overline{\epsilon} > \frac{1}{3}( \leq
\frac{1}{3})$ if $\int T^{\alpha}_{\alpha} \sqrt{-g} d^{3}x > 0 (\leq
0)$.

	Let us first consider the case $\overline{\epsilon} >
\frac{1}{3}$, $S<0$, where it can be shown that $R$ has a minimum value
at
\begin{equation}
R=R_{min} \equiv 2GM' \frac{(1+U)^{1+U}}{U^{U}}\;;\; U \equiv - S>0.
\end{equation}
Consequently, the solution does not cover the totality of space, since
it leaves out the region $0<R<R_{min}$.  Then, in order to be able to
match an interior static solution to this exterior solution, it must be
required that the radius of the interior configuration is larger than
$R_{min}$, thus implying that
\begin{equation}
R>R_{min} = \frac{2GM(1+U)^{1+U}}{c\;\;U^{U}}> 2GM,
\end{equation}
where we used $M=cM'$.  The value of $U$ can be calculated as a function
of $\omega$ for a given interior model using Eqs. (5.15) and (5.16).

	We now turn to the case $\overline{\epsilon} < \frac{1}{3}$,
$S>0$, which implies that
\begin{equation}
R \geq 0 \;;\;r \geq 2GM',
\end{equation}
and therefore no lower bound appears for $R$ from the above considerations of
the
exterior space-time alone, and further interior analysis is necessary.

	Besides their mathematical interest and heuristic value, the
solutions (3.6)-(3.9) provide a necessary tool for confronting with
reality the predictions of a very large class of viable theories, since
with the more refined observations of strong fields it is increasingly
important to use exact results.  In particular, the solutions of this
paper serve to investigate the trajectories of photons and particles
in the space exterior to static systems.  For instance, if the
configurations are spherically symmetric, then, from
\begin{equation}
\overline{U}^{\alpha} \overline{U}_{\alpha} =
\overline{U}^{\varphi}\overline{U}_{\varphi} + \overline{U}^{r}
\overline{U}_{r} +
\overline{U}^{4} \overline{U}_{4} = -1,
\end{equation}
\begin{equation}
\overline{U}^{\alpha} \equiv \frac{dX^{\alpha}}{d
\overline{\tau}};d\overline{\tau}^{2} = -\overline{g}_{\mu \nu} dX^{\mu}
dX^{\nu} = \phi d \tau^{2} ,
\end{equation}
we derived the following expressions for the orbits:
\begin{equation}
\varphi(r)-\varphi(r_{o}) = \pm \int^{r}_{r_{o}} \frac{L \;\;
dr}{\overline{R}^{2}[1+\overline{g}_{44}(\frac{1}{\overline{U}^{2}_{4}}+
\frac{L^{2}}{\overline{R}^{2}})]^{\frac{1}{2}}},
\end{equation}
\begin{equation}
t(r)-t(r_{0}) = \pm \int^{r}_{r_{o}}
\frac{dr}{\overline{g}_{44}[1+\overline{g}_{44}(\frac{1}{\overline{U}_{4}^{2}}+
\frac{L^{2}}{\overline{R}^{2}}]^\frac{1}{2}},
\end{equation}
where, without loss of generality, we have assumed that the motions occur
in the plane $\theta=\pi/2$, and
\begin{equation}
L \equiv \mid \frac{\overline{U}_{\varphi}}{\overline{U}_{4}} \mid  ,
\end{equation}
is the so-called angular momentum per mass or impact parameter.  Since
the field are independent of $t$ and $\varphi$, then the energy-momentum
conservation, $T^{\mu \nu}; \nu = 0$, yields
\begin{equation}
P_{\varphi} = m U_{\varphi} = \frac{m \overline{U}_{\varphi}}{\phi
^{\frac{1}{2}}}=const.; m=const.,
\end{equation}
\begin{equation}
P_{4}=mU_{4} + eA_{4} = \frac{m
\overline{U}_{4}}{\phi^{\frac{1}{2}}}+ eA_{4} = const.,
\end{equation}
from which we obtain that
\begin{equation}
L = \frac{\overline{L}}{1 - \frac{eA_{4} \overline{L}}{P_{\varphi}}};
\overline{L} = const.,
\end{equation}
and thus, for a neutral particle $L=\overline{L}$, and
$\overline{U}_{4} = const. \phi^{\frac{1}{2}}$.  Therefore, with the
metric (4.12), (4.13) and the expressions (5.22), (5.23), we are in a
position to extract the values of $\varphi (r)$ and $t(r)$, from
computer tabulations, and hence to obtain precise predictions for
relativistic celestial mechanics effects, like time delay, light
deflection, perihelium shift, etc.

\section{Magnetostatic Solutions}
\setcounter{equation}{0}
	Let us now assume that the metric and fields are static and
independent of a coordinate $X^{a}$.  Then we can generate magnetostatic
JBD like solutions starting from an Einstein's vacuum static metric,
$g'_{\mu \nu}$, in the following way (Appendix A):
\begin{equation}
- \overline{g}_{44} = - \frac{(-g'_{44})^{c}}{4a_{m}b_{m}}\left [ b_{m}
+ g'_{aa}(-g'_{44})^{1-c} \right]^{2} \equiv
\frac{(-g'_{44})^{c}}{\psi_{m}},
\end{equation}
\begin{equation}
\overline{g}_{aa} = \psi_{m} (-g'_{44})^{1-c} g'_{aa},
\end{equation}
\begin{equation}
\overline{g}_{ij} = \frac{(-g'_{44})^{1-c}}{\psi_{m}} g'_{ij} \;;\;i,j
\neq 4,a,
\end{equation}
\begin{equation}
A_{a} \equiv \pm \sqrt{a_{m}} \frac{[g'_{aa}(-g'_{44})^{1-c} -
b_{m}]}{g'_{aa}(-g'_{44})^{1-c} + b_{m}} + d_{m},
\end{equation}
\begin{equation}
\tilde{\Phi} = e\;\; \ell n (-g'_{44}) + \hbox{const.},
\end{equation}
where
\begin{equation}
\psi_{m} \equiv \frac{4a_{m}\; b_{m}}{[b_{m}+g'_{aa}(-g'_{44})^{1-c}]^{2}}.
\end{equation}
In order for $\overline{g}_{\mu \nu}$ to reduce to $g'_{\mu \nu}$ when
$c=1$ and $A_{a}=0$, the constants $a_{m}, b_{m} $ and $d_{m}$ should be
related as follows:
\begin{equation}
d_{m}= \pm \sqrt{a_{m}},
\end{equation}
\begin{equation}
b_{m} = 4a_{m}.
\end{equation}

	We can get explicit magnetostatic solutions by choosing as
background the axially symmetric Weyl metric (4.1)-(4.4).  In this case,
$X^{a}$ can be identified with the azymuthal angle $\varphi$, and we obtain
\begin{equation}
\overline{g}_{44} = - \Lambda^{2} \left [ 1- \frac{2
\beta}{r}\right]^{\delta c},
\end{equation}
\begin{equation}
\overline{g}_{rr} = \frac{\Lambda^{2}[1-\frac{2
\beta}{r}]^{\delta^{2}-\delta c-1}}{[1-\frac{2 \beta}{r} +
\frac{\beta^{2}}{r^{2}} sin^{2} \theta]^{\delta ^{2}-1}},
\end{equation}
\begin{equation}
\overline{g}_{\theta \theta} = \frac{\Lambda^{2}r^{2}[1-\frac{2
\beta}{r}]^{\delta^{2}- \delta c}}{[1-\frac{2 \beta}{r} +
\frac{\beta^2}{r^{2}} sin^{2} \theta]^{\delta ^{2}-1}},
\end{equation}
\begin{equation}
\overline{g}_{\varphi \varphi} = \frac{r^{2}[1-\frac{2 \beta}{r}]^{1-
\delta c}}{\Lambda^{2}}sin^{2} \theta,
\end{equation}
\begin{equation}
A_{\varphi} = \frac{Br^{2}[1-\frac{2\beta}{r}]^{1- \delta
c}}{\Lambda}sin^{2} \theta,
\end{equation}
\begin{equation}
\tilde{\Phi} = \delta e\;\; \ell n \left [ 1 - \frac{2 \beta}{r}\right
],
\end{equation}
where
\begin{equation}
\Lambda = B^{2}r^{2} \left [ 1- \frac{2 \beta}{r} \right ]^{1- \delta c}
sin^{2} \theta + 1,
\end{equation}
\begin{equation}
B \equiv \frac{\pm 1}{2 \sqrt{a_{m}}}.
\end{equation}

When $\delta$ = 1, $c$=1, and $\beta$=0, the above solutions reduce to the
well known magnetized Melvin
solutions$^{14}$, which represents a cylindrically symmetric bundle of
pure magnetic flux.  If $\beta \neq 0$ we have a Schwarzschild black
hole embedded in the magnetic flux$^{15}$.  If, furthermore, $\delta \neq 1$
and $c \neq 1$, the configuration is not spherically symmetric, and a scalar
field is also present.

\section{Acknowledgements}

I am grateful to the Department of Physics of The Pennsylvania State
University for their hospitality and support, and the University of
Puerto Rico at Humacao for their financial support.

\renewcommand{\theequation}{\Alph{section}.\arabic{equation}}

\newpage
\section*{Appendix A}
\setcounter{section}{1}
\setcounter{equation}{0}

	Suppose that the JBD like field equations are independent
of the coordinate $X^{a}$, and that the metric and electromagnetic
vector have the form
\begin{equation}
\overline{g}_{\mu \nu} = \left(
\begin{array}{cl}
                  \overline {g}_{aa} & 0 \nonumber \\
                                  0  & \overline{g}_{ij}
\end{array}
\right),
\end{equation}
\begin{equation}
A_{\mu} = A_{a} \delta_{\mu a}.
\end{equation}
Then, with the following transformation of the metric
\begin{equation}
\overline{g}_{aa} = g'_{aa} \psi \Omega,
\end{equation}
\begin{equation}
\overline{g}_{ij} = \frac{g'_{ij}}{\psi \Omega},
\end{equation}
we obtain, from the JBD like equations,
\begin{equation}
\overline{G}_{ij} - 8 \pi G \overline{T}_{ij} - (\omega_{0} +
\frac{3}{2}) [\tilde{\Phi}_{,i} \tilde{\Phi}_{,j}-\frac{1}{2}
\overline{g}_{ij} \tilde{\Phi}^{,k} \tilde{\Phi}_{,k}] = G'_{ij} -
\frac{C_{ijlk}}{2}[A^{lk}+B^{lk}]=0,
\end{equation}
where
\begin{equation}
C_{ijlk} \equiv g'_{il} g'_{jk} - \frac{1}{2} g'_{ij} g'_{lk},
\end{equation}
\begin{equation}
A^{lk} \equiv (\ell n g''_{aa})^{,l}(\ell n \psi)^{,k} +
(\ell n g''_{aa})^{,k} (\ell n \psi)^{,l} + (\ell n \psi)^{,l}(\ell n
\psi)^{,k} + \frac{4A^{,l}_{a}A^{,k}_{a}}{g''_{aa}\psi},
\end{equation}
\begin{equation}
B^{lk} \equiv (\ell n g'_{aa})^{,l}(\ell n \Omega)^{,k} + (\ell n
g'_{aa})^{,k}(\ell n \Omega)^{,l} + (\ell n \Omega)^{,l} (\ell n
\Omega)^{,k} + (2 \omega_{0} + 3) \tilde \Phi^{,l} \tilde{\Phi}^{,k} ,
\end{equation}
\begin{equation}
g''_{aa} \equiv g'_{aa} \Omega.
\end{equation}
Furthermore,
\begin{equation}
\overline{R}^{a}_{a} = \psi \Omega R'^{a}_{a} - \frac{1}{2} \frac{\psi
\Omega[\sqrt{-g'}(\ell n \psi \Omega)^{,k}]_{,k}}{\sqrt{-g'}} =
\frac{A_{a}^{,k}A_{a,k}}{g'_{aa}} ,
\end{equation}
\begin{equation}
F^{a \nu}\,_{; \nu} = \frac{- \psi \Omega}{\sqrt{-g'}}\left [
\frac{\sqrt{-g'}A_{a}\;^{,k}}{g''_{aa}\psi}\right ]_{,k}=0 ,
\end{equation}
\begin{equation}
\tilde{\Phi}^{; \alpha}_{; \alpha} = \frac{\psi \Omega}{\sqrt{-g'}}\left
[ \sqrt{-g'}\Phi^{,k} \right ]_{,k} = 0  .
\end{equation}
Note that in the right hand side of the equations the indexes are raised
with $g'^{ij}$.

	We are looking for functions $\psi$ and $\Omega$ such that
$A^{lk} =0, B^{lk}=0$, and, thereby, $G'_{ij}=0$.  The inspection of the
expressions (A.7) and (A.8) suggest that we work with the following
ansatz:
\begin{equation}
A_{a}=A_{a}(g'_{aa})\; , \; \psi = \psi (g'_{aa}) ,
\end{equation}
\begin{equation}
\tilde{\Phi} = \tilde{\Phi}(g'_{aa}) \; ; \; \Omega = \Omega(g'_{aa}).
\end{equation}
The hipothesis (A.13) and (A.14) are equivalent to the assumptions
(3.3)-(3.5), and hence these are solutions where the surfaces $u \equiv
g^{'}_{aa} = const.$ are equipotentials of $A_{a}, \overline{g}_{aa}$
and $\tilde{\phi}$.  Thus, for instance, for static systems, $a=4$, the
surfaces of constant gravitational and electric potentials coincide,
which is to be expected in astrophysical systems in equilibrium.

	Aside from the assumptions (A.13) and (A.14) the functions
$\psi$ and $\Omega$ are arbitrary and can be chosen to satisfy the
following relations
\begin{equation}
\frac{d \ell n \psi}{d \ell n g''_{aa}}= -1 \pm \sqrt{1-4 g''_{aa} \psi
k^{2}_{1}},
\end{equation}
\begin{equation}
\frac{d \ell n \Omega}{d \ell n g'_{aa}} = -1 \pm \sqrt{1-k^{2}_{2}},
\end{equation}
where
\begin{equation}
k^{2}_{1} \equiv \left [ \frac{1}{\psi} \frac{dA_{a}}{d g''_{aa}} \right
]^{2}  ,
\end{equation}
\begin{equation}
k^{2}_{2} \equiv (2 \omega_{0} +3) \left ( \frac{d \tilde \Phi}{d \ell n
g'_{aa}} \right )^{2}   ,
\end{equation}
which can be shown to guarantee that $A^{\ell k}$ and $B^{\ell k}$
vanish.  Then (A.5) implies that $G^{'}_{ij}=0$, and as a consequence of
the Bianchi identities, $G^{'\mu \nu}_{; \nu} = 0$, we also have
$G^{'}_{aa}=0$, or $G^{'}_{\mu \nu} =0$.

On the other hand, from Einstein's vacuum equations, $G^{'}_{\mu \nu} =
0$, we get
\begin{equation}
R'^{a}_{a} = - \frac{1}{2\sqrt{-g'}}\left[ \sqrt{-g'}(\ell n
g'_{aa})^{,k} \right ]_{,k} = 0,
\end{equation}
which together with Eqs. (A.11), (A.12), (A.16) and (A.17) imply, if $g'_{ij}
\geq 0$, that
\begin{equation}
k_{1} = const.,
\end{equation}
\begin{equation}
k_{2}=const.
\end{equation}
Then from Eqs. (A.15)-(A.18), we obtain
\begin{equation}
\overline{g}_{aa} = g''_{aa}\psi = \frac{4abg''_{aa}}{[b+g''_{aa}]^{2}} ;
k_{1}^{2} \equiv \frac{1}{4a}, b = const.,
\end{equation}
\begin{equation}
A_{a} = \pm \frac{\sqrt{a}[b-g''_{aa}]}{[b+g''_{aa}]^{2}} + const.,
\end{equation}
\begin{equation}
\Omega = g'_{aa}\;^{c-1} ; c \equiv \pm \sqrt{1-k^{2}_{2}},
\end{equation}
\begin{equation}
\ell n \tilde{\Phi} =\frac{\pm k_{2} \ell n g'_{aa}}{\sqrt{2
\omega_{0}+3}}+ const.
\end{equation}
We can also verify that Eq. (A.10) is satisfied by the above equations.

	Equations (A.22)-(A.25) and Eq. (A.4) can be converted into Eqs.
(3.6)-(3.9) by identifying `a' with `4'.  If, on the other hand, $x^{a}$ is
a space-like
coordinate, then $A_{a}$ represents a magnetic field.

	If the fields are also independent of another coordinate $x^{b}$,
then we can consider instead the conformal transformation
\begin{equation}
\overline{g}_{aa} = \frac{g'_{aa} \psi}{\Omega},
\end{equation}
\begin{equation}
\overline{g}_{bb} = \frac{g'_{bb}\Omega}{\psi},
\end{equation}
\begin{equation}
\overline{g}_{ij} = \frac{g'_{ij}}{\psi \Omega}\;;\; i,j \neq a,b.
\end{equation}
and then Eqs. (A.5) are derived with the following changes in $A^{\ell
k}$ and $B^{\ell k}$; now in $A^{\ell k}$, $g''_{aa}$ means
\begin{equation}
g''_{aa} \equiv \frac{g'_{aa}}{\Omega};
\end{equation}
and in $B^{\ell k}$, $g'_{aa}$ is substituted by $g'_{bb}$.
Consequently, following the same routine that led to Eqs. (A.22) -
(A.25), we will again obtain Eqs. (A.22) and (A. 23), but now
\begin{equation}
\Omega = {g'_{bb}}^{c-1},
\end{equation}
and
\begin{equation}
\ell n \tilde{\Phi} = \frac{\pm k_{2}}{\sqrt{2\omega_{0}+3}} \ell n
g'_{bb} + const.
\end{equation}
Choosing $b=4$ gives rise to the magnetostatic solutions of Section VI.
If, on the other hand, $x_{b}$ is space-like, then we would get
electrostatic $(a=4)$ or magnetic solutions $(a \neq 4)$ that were not
studied in this paper.

\newpage
\section*{Appendix B}
\setcounter{section}{2}
\setcounter{equation}{0}

	We see from Eqs. (4.5), (4.9) and (4.10) that the surfaces with
r = const. are equipotentials for the gravitational, electric, and scalar
fields.  It is easy to calculate the proper distance on these surfaces
along the circumferences $\theta$ = const., and we get, using Eq. (4.8)
\begin{equation}
C_{\varphi}{(\theta)} = \int^{2 \pi}_{0} \sqrt{\overline{g}_{\varphi
\varphi}} d\varphi = \frac{2 \pi r[1-\frac{2
\beta}{r}]^{\frac{1-\delta c}{2}}}{\sqrt{\psi}} sin \theta \equiv 2 \pi
\overline{R} sin \theta.
\end{equation}
On the other hand, the proper distance around the curves $\varphi =
\hbox{const.}$ are given by the integral
\begin{equation}
C_{\theta} = 2 \int^{\pi}_{0} \sqrt{\overline{g}_{\theta \theta}}d
\theta = 2 \overline{R} \int^{\pi}_{0} \left ( 1+
\frac{\beta^{2}sin^{2}\theta}{r^{2}[1-\frac{2 \beta}{r}]}\right
)^{\frac{1- \delta^{2}}{2}} d \theta
\end{equation}
where we used Eq. (4.7).  We can see that since $r>2 \beta$,
\begin{equation}
C_{\theta} > C_{\varphi}(90^{\circ}) = 2 \pi \overline{R}\; ; \; 0<\delta<1,
\end{equation}
\begin{equation}
C_{\theta} < C_{\varphi}(90^{\circ}) \;\; , \;\; \delta > 1,
\end{equation}
and therefore the equipotential surfaces have shapes like prolates
($0<\delta<1$) or oblates ($\delta > 1$).

\newpage
\section*{Appendix C.  The Reissner-Nordstr$\ddot{o}$m Solution.}
\setcounter{section}{3}
\setcounter{equation}{0}

	If $c=1$, one finds from Eq. (4.12) that
\begin{equation}
r = \overline{R} - \frac{2 \beta}{b-1}  ,
\end{equation}
and then Eq. (4.13) becomes
\begin{equation}
- \overline{g}_{44} = \overline{g}_{rr}^{-1} = 1 - \frac{2
\beta \left (\frac{b+1}{b-1}\right )}{\overline{R}} + \frac{4
\beta^{2}b}{(b-1)^{2} \overline{R}^{2}}
\end{equation}
On the other hand, recalling Eqs. (3.18) and (3.19), and since $c=1,
\beta = GM'$, we
find that
\begin{equation}
- \overline{g}_{44} = \overline{g}_{rr}^{-1} = 1 -
\frac{2GM}{\overline{R}}+k^{2} \frac{Q^{2}}{\overline{R}^{2}},
\end{equation}
which together with Eq. (4.12) is the standard form for the
Reissner-Nordstr$\ddot{o}$m metric.  Furthermore, Eq. (4.9) with $\delta =
c = 1$ gives the familiar electrostatic potential:
\begin{equation}
A_{4} = \frac{-kQ}{\overline{R}}
\end{equation}

\newpage
\begin{enumerate}
\item T. Singh and T. Singh, International Journal of Modern
Physics A, Vol. 2, \#3, 645 (1987).
\item P. G. Bergmann, Int. J. Theor. Phys. \underline{1}, 25
(1968).
\item R. V. Wagoner, Phys. Rev. D\underline{1}, 3209 (1970).
\item K. Nordtvedt, Astrophysics J. \underline{161}, 1059 (1970).
\item C. H. Brans and R. H. Dicke, Phys. Rev. \underline{124}, 925
(1961).
\item A. Einstein and W. Mayer, Akad-Wiss Berlin, K1, Phys. Math.
\underline{22}, 541 (1931).
\item P. Jordan, Ann. Physik, p. 219 (1947).
\item Y. Thiry, Acad. Sci. \underline{226}, 216 (1948).
\item C. W. Misner, K. S. Thorne and J. A. Wheeler, Gravitation
(W. H. Freeman, San Francisco, 1973).
\item B. M. Barker, Astrophy. J. \underline{219}, 5 (1978).
\item J. Schwinger, Particles, Sources and Field (Addison-Wesley,
Reading, MA, Vol. 1, 1970).
\item R. H. Dicke, Phys. Rev. \underline{125}, 2163 (1962).
\item B.H. Voorhees, Phys. Rev. D\underline{2}, 2119 (1970).
\item M. A. Melvin, Phys. Letts. \underline{8}, 65 (1964).
\item J. F. Ernst and W. J. Wild, J. Math. Phys. \underline{17},
182 (1976).
\end{enumerate}
\end{document}